\documentclass[aps, prl, twocolumn, preprintnumbers, superscriptaddress, showpacs, amssymb]{revtex4-1}

\usepackage{graphicx}
\usepackage{dcolumn}
\usepackage{bm}

\begin{document}

\preprint{Ver. 6sub}¡¡

\title{Anomalous peak effect in iron-based superconductors Ba$_{1-x}$K$_x$Fe$_2$As$_2$ ($x \approx$ 0.69 and 0.76) for magnetic-field directions close to the $ab$ plane and its possible relation to the spin paramagnetic effect}


\author{Taichi Terashima}
\author{Naoki Kikugawa}
\affiliation{National Institute for Materials Science, Tsukuba, Ibaraki 305-0003, Japan}
\author{Andhika Kiswandhi}
\altaffiliation[Present address: ]{Department of Chemistry, Graduate School of Science, Kyoto University, Kyoto 606-8502, Japan}
\author{Eun-Sang Choi}
\affiliation{National High Magnetic Field Laboratory, Florida State University, Tallahassee, FL 32310, USA}
\author{Kunihiro Kihou}
\author{Shigeyuki Ishida}
\author{Chul-Ho Lee}
\author{Akira Iyo}
\author{Hiroshi Eisaki}
\affiliation{National Institute of Advanced Industrial Science and Technology (AIST), Tsukuba, Ibaraki 305-8568, Japan}
\author{Shinya Uji}
\affiliation{National Institute for Materials Science, Tsukuba, Ibaraki 305-0003, Japan}


\date{\today}
\begin{abstract}
We report magnetic torque measurements on iron-pnictide superconductors Ba$_{1-x}$K$_x$Fe$_2$As$_2$ ($x \approx$ 0.69 and 0.76) up to an applied field of $B_a$ = 45 T.
The peak effect is observed in torque-vs-field curves below the irreversibility field.
It is enhanced and becomes asymmetric as the field is tilted from the $c$ axis.
For field directions close to the $ab$ plane, increasing- and decreasing-field curves peak at markedly different fields, and exhibit a sharp jump, suggestive of a first-order phase transition, on the high- and the low-field side of the peak, respectively.
Complicated history dependence of the torque is observed in the peak-effect region.
We construct and discuss the temperature ($T$)--applied-magnetic-field ($B_a$) phase diagram.
Since the upper critical field for the $ab$-plane direction is comparable to the Pauli limit, we also consider possible influence of the spin paramagnetic effect on the anomalous peak effect.
\end{abstract}


\maketitle



\newcommand{\ud}{\mathrm{d}}
\def\degree{\kern-.2em\r{}\kern-.3em}

\section{Introduction}
The critical current in type-II superconductors often shows an anomalous peak just before it becomes zero at the upper critical field $B_{c2}$.
This `peak effect' has been known since the early 1960's and has attracted continuing attention \cite{Berlincourt61PRL, Pippard69PhilMag}.
Its mechanism however still remains unresolved.
Not only will its elucidation deepen our understanding of the vortex-matter physics, it may also be of technological importance: it might open a new avenue to improve or tailor the critical current.

One of plausible explanations associates the peak effect with an order-disorder transition of the vortex lattice \cite{Banerjee01PhysicaC, Giamarchi02Book}.
The temperature ($T$)--applied-magnetic-field ($B_a$) phase diagram of ideal type-II superconductors consists of the Meissner and mixed states.
However, in real materials, the perfect Abrikosov vortex lattice does not exist, and the mixed state is subdivided into different vortex states.
A quasi-long-range-ordered Bragg glass occupies a low-$T$ low-$B_a$ part of the mixed state in weak-pinning superconductors \cite{Giamarchi95PRB, Klein01Nature}.
The Bragg glass melts into a vortex liquid as the temperature increases.
On the other hand, increasing magnetic field is equivalent to increasing pinning strength.
As the field is increased at low temperatures, the Bragg glass thus disorders at a certain field to better adapt to the random pinning environment, resulting in a larger critical current.
Although the nature of the disordered phase is still controversial, it is widely believed that this order-disorder transition underlies the peak effect \cite{Banerjee01PhysicaC, Giamarchi02Book}.
Experimental evidence has accumulated, especially in low-$T_c$ materials:
Magnetic measurements showed anomalous field- or temperature-history dependence in the peak effect region \cite{Roy00PRB, Ravikumar00PRB, Angst03PRB}.
The coexistence of two phases with differing critical currents in the peak-effect region were directly seen by scanning Hall-probe microscopy \cite{Marchevsky01Nature}.
Small angle neutron scattering revealed disordering of the vortex lattice near the peak-effect region \cite{Gammel98PRL, Ling01PRL}.

In this article, we report magnetic torque measurements on iron-pnictide superconductors Ba$_{1-x}$K$_x$Fe$_2$As$_2$ ($x \approx$ 0.69 and 0.76).
Compounds with those compositions are in the over-doped regime and exhibit no magnetic transition.
We find an anomalous peak effect for magnetic field directions close to the $ab$ plane:
The peak positions of torque-vs-field curves differ significantly between increasing- and decreasing-field sweeps.
Further, increasing- and decreasing-field curves exhibit a sharp jump, suggestive of a first-order transition, on the high- and low-field side of the peak, respectively.
We construct $T$-$B_a$ phase diagrams and discuss the experimental results.

\section{Experimental details}

\begin{figure}
\includegraphics[width=8.6cm]{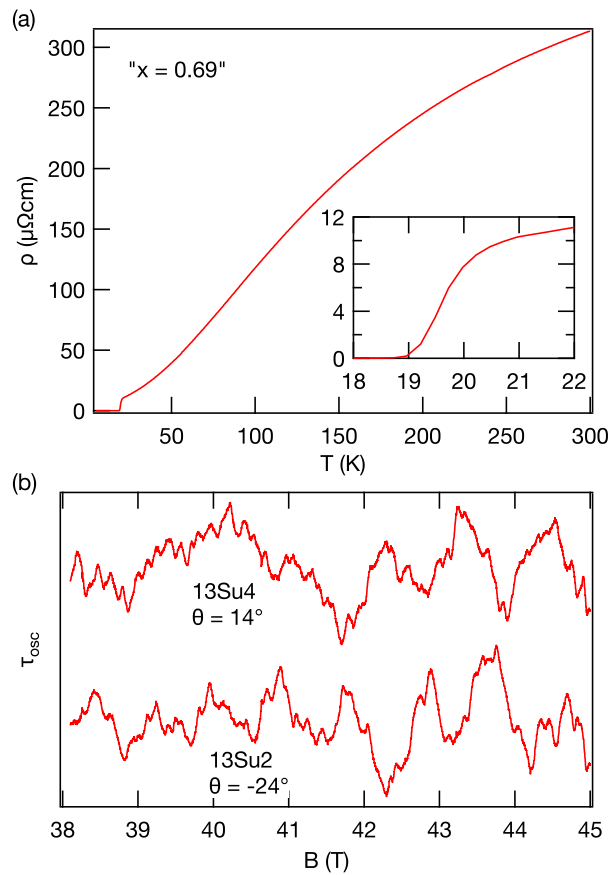}
\caption{\label{RvsT} (a) Resistivity vs temperature for a crystal from the $x \approx 0.69$ batch.  The inset is a blow-up of the transition region.  (b)  De Haas-van Alphen oscillations in the magnetic torque for samples 13Su4 and 13Su2 at $T$ = 0.41 K.  The dominant frequency is $F \approx 1.7$ and 1.9 kT, respectively.  A polynomial smooth background has been subtracted from each raw data.}   
\end{figure}

Single crystals of Ba$_{1-x}$K$_x$Fe$_2$As$_2$ ($x \approx 0.69$ and 0.76) were synthesized by a KAs self-flux method \cite{Kihou10JPSJ, Kihou16JPSJ}.
To determine the composition $x$, energy-dispersive X-ray analyses were applied to crystals from two growth batches.
For each batch, the composition varied from crystal to crystal by $\pm$ a few percent.
The compositions $x$ of 0.69 and 0.76 are the average values.
Resistivity $R$ measurements on one crystal from the $x \approx 0.69$ growth batch showed the superconducting transition temperature of $T_c$ = 19.6 K with the transition width of 1.1 K [Fig. \ref{RvsT}(a)].
The resistivity ratio defined as $R(300 \mathrm{K})/R(21 \mathrm{K})$ was 30, indicating high quality of the crystal.

For magnetic torque measurements, small pieces with typical dimensions of (50--100 $\mu$m)$^2 \times$ (a few tens of $\mu$m) were prepared by cleaving crystals along $<$100$>$ axes:
sample 13Su4 ($19 \mathrm{K} < T_c < 21 \mathrm{K}$) is from the $x \approx 0.69$ batch, and samples 13Su2 ($15 \mathrm{K} < T_c < 16 \mathrm{K}$), 13Sp4 ($19 \mathrm{K} < T_c < 21 \mathrm{K}$), and 14Sp4 are from the $x \approx 0.76$ batch.
$T_c$ was estimated from the temperature dependence of the torque hysteresis curves [Figs. \ref{13Su4}(b), \ref{13Su2}(b), and \ref{13Sp4}(a)] (sample 14Sp4 was measured only at the base temperature).
The 45-T hybrid magnet or a 35-T resistive magnet was used with a $^3$He refrigerator at the NHMFL in Tallahassee.
The magnetic torque $\bm{\tau}=\bm{M}\times\bm{B_a}$ was measured with a piezoresistive microcantilever \cite{Ohmichi02RSI}.
The angle $\theta$ of the applied field $\bm{B_a}$ was measured from the $c$ axis.
De Haas-van Alphen oscillations were observed in samples 13Su4, 13Su2, and 13Sp4 for field directions near the $c$ axis [see Fig. \ref{RvsT}(b) for data for 13Su4 and 13Su2], which confirms the high quality of the crystals.

\section{Results}

\begin{figure}
\includegraphics[width=8.6cm]{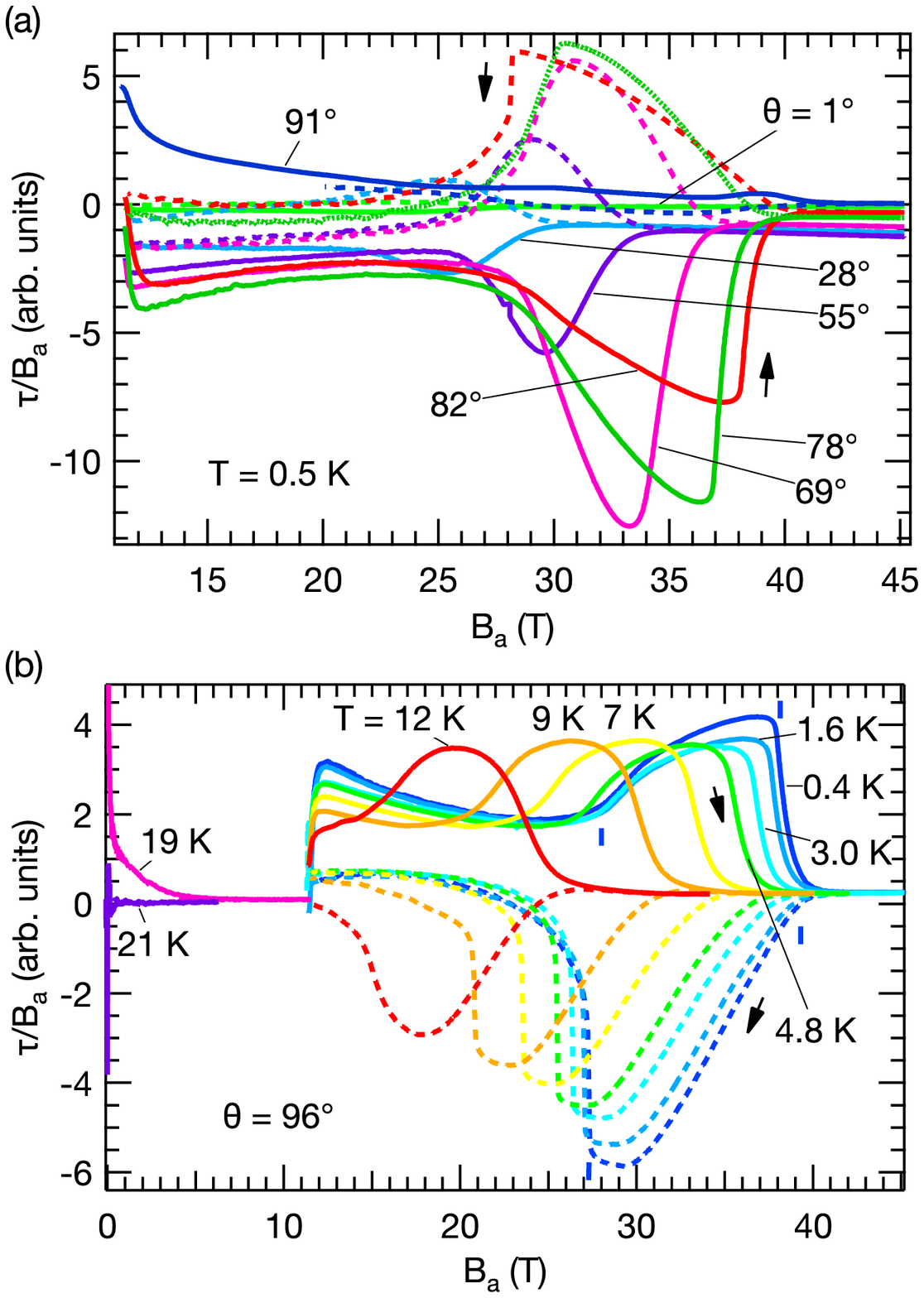}
\caption{\label{13Su4}Hysteresis loops of sample 13Su4 ($19 \mathrm{K} < T_c < 21 \mathrm{K}$) for various field directions (a) and for various temperatures (b).  Magnetic torque divided by applied field is shown as a function of applied field.  The solid and broken curves show increasing- and decreasing-field ones, respectively, as indicated by the arrows.  The vertical bars in (b) indicate the four characteristic fields $B_{1}^{+(-)}$ and $B_{2}^{+(-)}$ (see text for the definitions) at $T$ = 0.4 K.}   
\end{figure}

\begin{figure}[!]
\includegraphics[width=8.6cm]{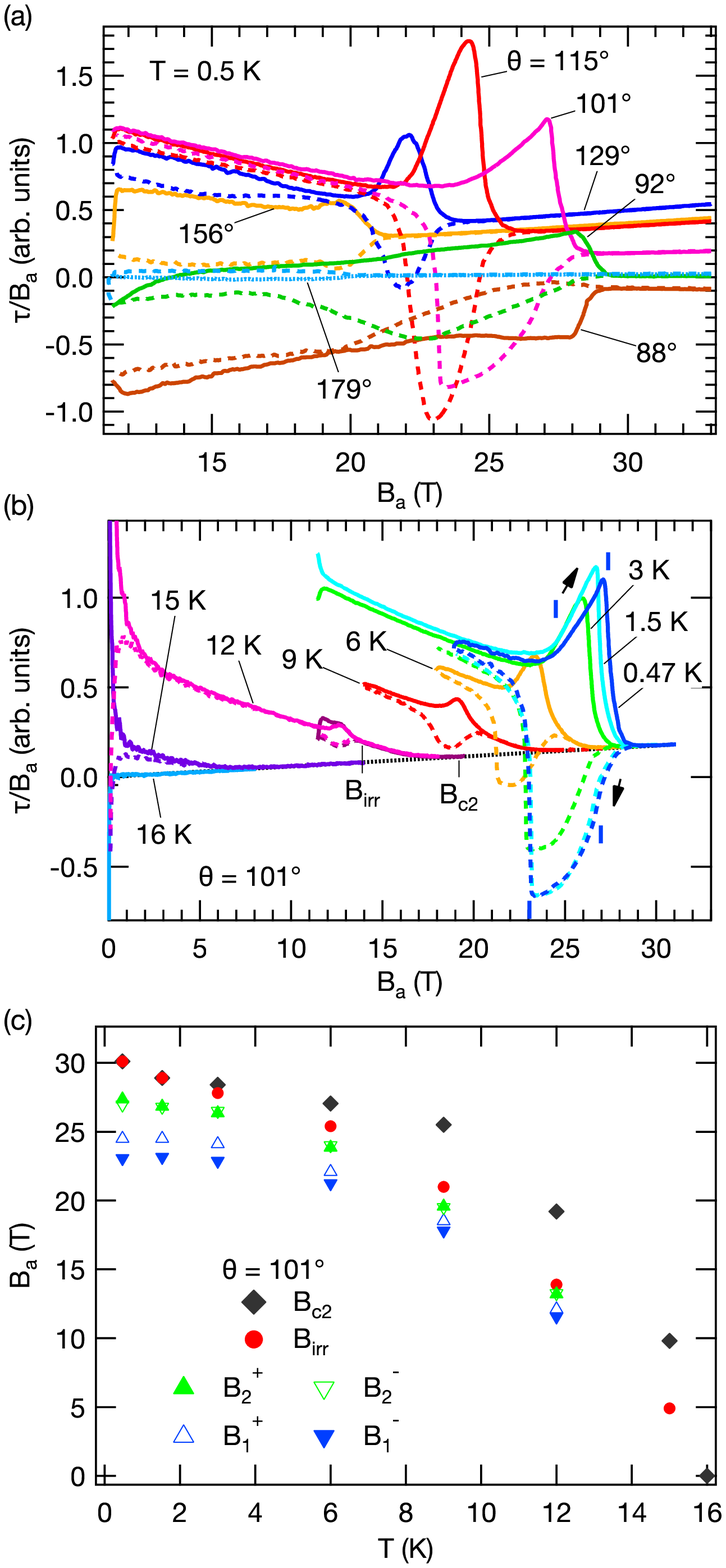}
\caption{\label{13Su2} (a) and (b)  Hysteresis loops of sample 13Su2 ($15 \mathrm{K} < T_c < 16 \mathrm{K}$) for various field directions (a) and for various temperatures (b).  Magnetic torque divided by applied field is shown as a function of applied field.  The solid and broken curves show increasing- and decreasing-field ones, respectively, as indicated by the arrows in (b).  The vertical bars in (b) indicate the four characteristic fields $B_{1}^{+(-)}$ and $B_{2}^{+(-)}$ (see text for the definitions) at $T$ = 0.47 K.  (c)  Phase diagram derived from the data in (b).}   
\end{figure}

\begin{figure}
\includegraphics[width=8.6cm]{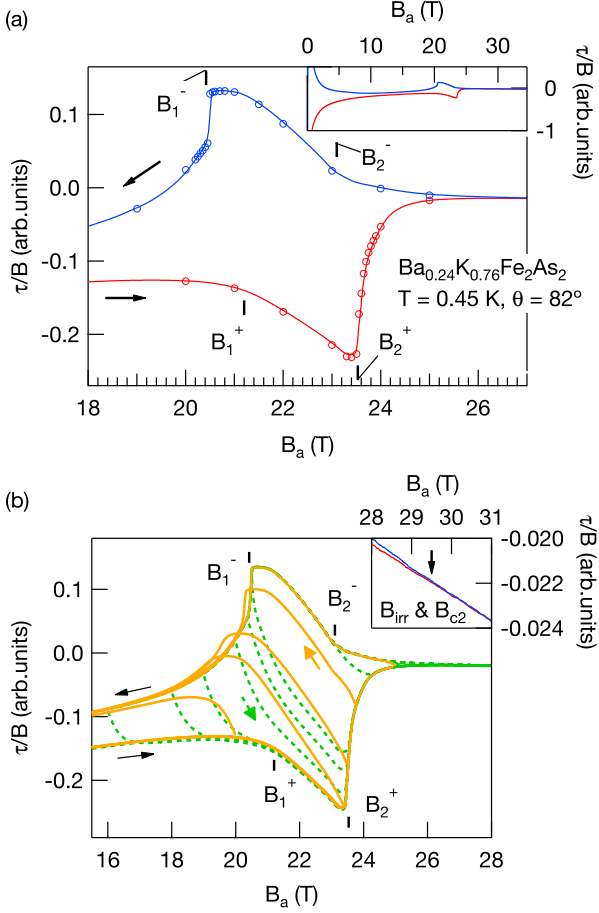}
\caption{\label{at82deg}Hysteresis loops and history dependence of sample 14Sp4 at $\theta$ = 82$^{\circ}$ and $T$ = 0.45 K.  (a) Enlarged view of the peak effect region.  The field sweep direction is indicated by arrows.  The solid line was obtained from a continuous field sweep, while circles by stopping field at some value and waiting for about one minute to see effects of relaxation.  See text for the definitions of the characteristic fields $B_{1}^{+(-)}$ and $B_{2}^{+(-)}$.  The full hysteresis curve is shown in the inset.  (b) Minor hysteresis loops showing history effects.  Solid curves are obtained by increasing field from 0 T to some field and then decreasing field, while broken ones by decreasing field from 33 T to some field and then increasing field, as indicated by arrows.  The inset is an enlarged view of a region near the irreversibility field and upper critical field.}   
\end{figure}

\begin{figure}
\includegraphics[width=8.6cm]{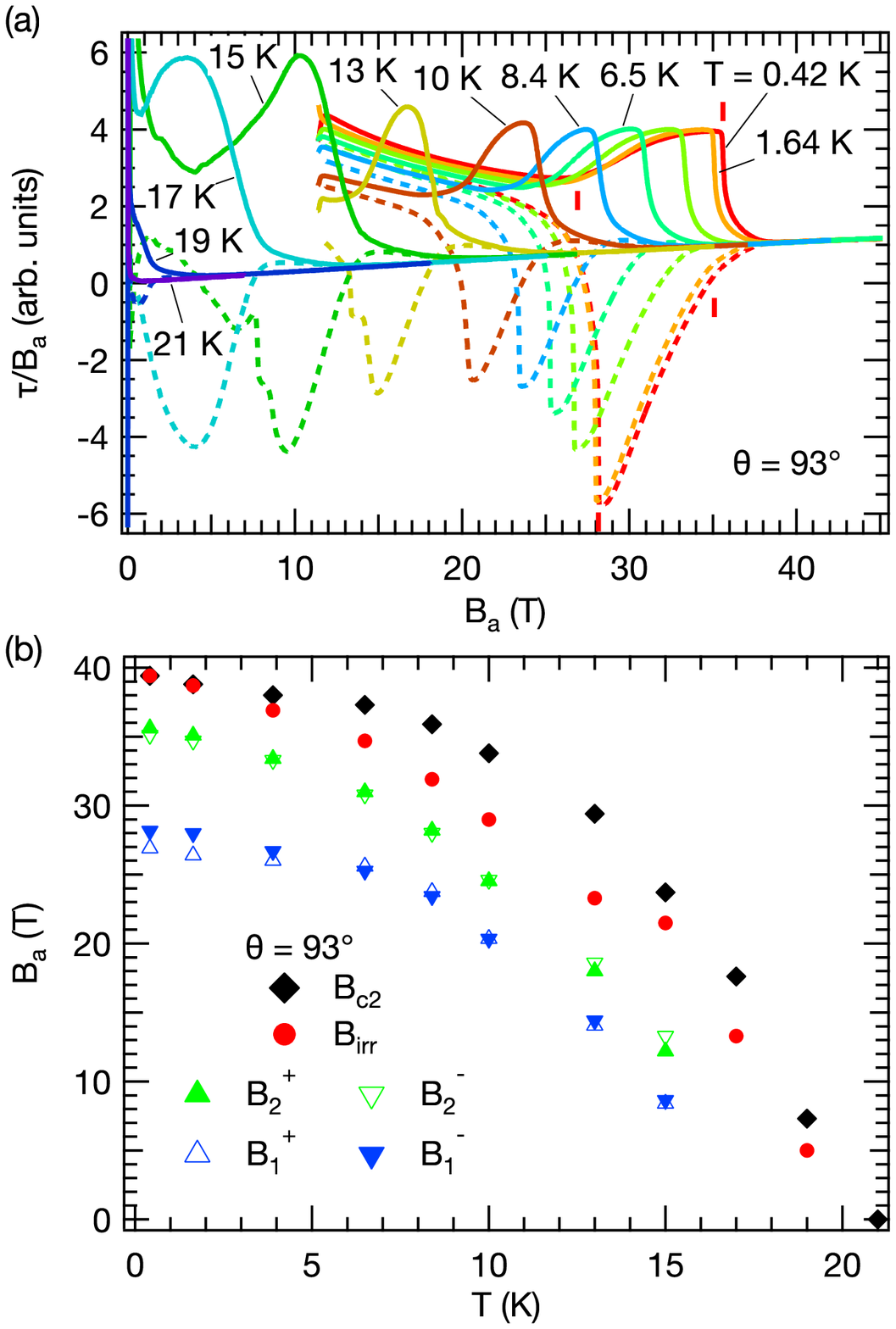}
\caption{\label{13Sp4} (a)  Hysteresis loops of sample 13Sp4 ($19 \mathrm{K} < T_c < 21 \mathrm{K}$) at $\theta$ = 93$^{\circ}$ for various temperatures.  Magnetic torque divided by applied field is shown as a function of applied field.  The solid and broken curves show increasing- and decreasing-field ones, respectively.  The vertical bars indicate the four characteristic fields $B_{1}^{+(-)}$ and $B_{2}^{+(-)}$ (see text for the definitions) at $T$ = 0.42 K.  (b)  Phase diagram derived from the data in (a).}   
\end{figure}

Figures~\ref{13Su4} (a) and \ref{13Su2}(a) show magnetic hysteresis loops of samples 13Su4 and 13Su2, respectively, at $T$ = 0.5 K for various field directions.
The vertical axis is the magnetic torque divided by the applied field, $\tau/B_a$, which corresponds to the magnetization normal to the field.
Since the hybrid magnet was used, the field was cycled between 11.5 and 45.1 T.
The difference in the torque $\Delta\tau$ between increasing- and decreasing-field curves at a given field is a measure of the critical current, or the pinning force, at that field.
The peak effect hence manifests itself as enhancement of $\Delta\tau$ just before the two curves merge at the irreversibility field $B_{irr}$.
The figures indicate that the peak effect becomes more pronounced as the field is tilted from the $c$ axis towards the $ab$ plane with increasing $\theta$.
The peak becomes asymmetric at large angles: an increasing- and a decreasing-field curve peak at markedly different fields.
An increasing-field curve shows a sharper slope on the high-field side of the peak while a decreasing-field one shows a sharper slope on the low-field side.

Figure~\ref{at82deg} shows hysteresis loops of sample 14Sp4 at $\theta$ = 82$^{\circ}$.
The inset of (a) shows a full hysteresis curve, while the main panel an enlarged view of the peak-effect region.
The solid line shows results of a continuous field sweep.
We define four characteristic fields based on d$(\tau /B)$/d$B$ ($B_1^-$ and $B_2^+$) and d$^2(\tau /B)$/d$B^2$ ($B_1^+$ and $B_2^-$).
The shape of the hysteresis loop in the peak-effect region is very different from roughly symmetric shapes observed in previous magnetic measurements on CeRu$_2$, NbSe$_2$, and MgB$_2$ \cite{Roy00PRB, Ravikumar00PRB, Angst03PRB}.
Further, those previous works did not observe features like the sharp changes of $(\tau /B)$ at $B_1^-$ and $B_2^+$, which suggest the existence of first-order phase transitions at these fields.

In order to verify that these anomalies are not artifacts caused by field sweeping, we have taken relaxation data at fields indicated by hollow circles.
The circles show $(\tau /B)$ measured after one-minute relaxation at the respective fields.
Clearly, the relaxation effects are negligible, and the anomalies at $B_1^-$ and $B_2^+$ can be seen in the relaxed torque.
The behavior of $(\tau /B)$ at $B_1^-$ indicates that the pinning is weaker below the transition field, while the behavior at $B_2^+$ indicates that the pinning is weaker above the transition field.

Figure~\ref{at82deg}(b) shows various minor hysteresis loops.
The curve branching off from the increasing-field curve at a field below $B_1^+$ undershoot the decreasing-field curve of the full loop, while those branching off at fields above $B_1^+$ overshoot.
Also, curves branching off from the decreasing-field curve at low fields go slightly below the increasing-field curve of the full loop.
These observations suggest complicated phase coexistence due to the first-order phase transitions.
Note that the branched-off curves traverse a large field difference to approach the opposite side of the full loop.
The curve branching off from the increasing-field curve at $B_a$ = 23.4 T just below $B_2^+$ does not reach the decreasing-field curve until 19.5 T, for example.

Figures~\ref{13Su4}(b), \ref{13Su2}(b), and \ref{13Sp4}(a) show temperature variation of hysteresis loops for a field direction near $\theta$ = 90$^{\circ}$.
As the temperature is raised, the anomalies at $B_1^-$ and $B_2^+$ becomes less sharp, and the increasing- and decreasing-curves become more symmetric.
It is also interesting to note that the irreversibility field becomes distinct from the upper critical field at elevated temperatures [see e.g. the $T$ = 12 K curve in Fig.~\ref{13Su2}(b)].
The derived $T$-$B_a$ phase diagrams are shown in Figs.~\ref{13Su2}(c) and \ref{13Sp4}(b) [For sample 13Su4 in Fig.~\ref{13Su4}(b), swept field ranges were insufficient to determine $B_{c2}$ and $B_{irr}$].

\section{discussion}

\begin{figure}
\includegraphics[width=8.6cm]{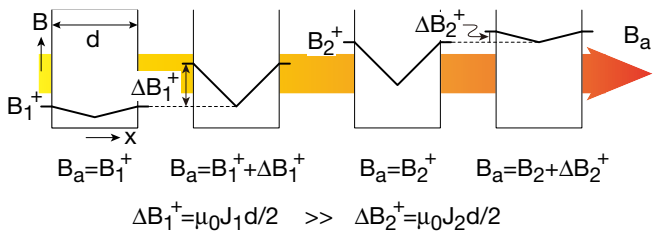}
\caption{\label{Bean}Explanation of the asymmetric hysteresis loops within the Bean model \cite{BEAN64RMP}, which assumes that the field gradient develops in a superconducting sample according to d$B$/d$x$ = $\mu_0J_c$ where $J_c$ is a critical current density (the current direction is normal to $B$ and $x$).  We assume that magnetic fields are applied parallel to the surface of a slab of a superconductor with a thickness $d$.  Each of the four plots shows the magnitude of a local field inside the slab at a given $B_a$ as a function of $x$, which is along the direction of the thickness $d$.   From left to right, the applied field $B_a$ is increased.  $J_c$ = $J_0$, $J_1$, and $J_2$ for $B_a < B_1^+$, $B_1^+ < B_a < B_2^+$, and $B_2^+ < B_a$, respectively, and $J_1 \gg J_0$, $J_2$.  When $B_1^+$ is crossed from below to above, the field gradient inside the slab changes only gradually from the surface, and hence the width of the transition region $\Delta B_1^+$ is large.  On the other hand, when $B_2^+$ is crossed, the critical current collapses and can no longer sustain the existing field gradient, giving rise to a quick change in the magnetization.  The transition width $\Delta B_2^+$ is much smaller.}   
\end{figure}

Let us first assume that the sharp anomalies at $B_1^-$ and $B_2^+$ are due to first-order phase transitions of the vortex matter and see how well we can explain the observed anomalous peak effect.
The behavior of $(\tau /B)$ at $B_1^-$ indicates that the pinning is weaker below $B_1^-$, while the behavior at $B_2^+$ indicates that the pinning is weaker above $B_2^+$ [see Fig. \ref{at82deg}(a)].
Therefore the two anomalies at $B_1^-$ and $B_2^+$ cannot be attributed to a single first-order transition: there are two separate phase transitions.
We assume that the counterpart of $B_1^-$ is $B_1^+$ and that that of $B_2^+$ is $B_2^-$.
Note however that the current definition of $B_1^+$ and $B_2^-$ based on d$^2(\tau /B)$/d$B^2$ needs to be improved since an unphysical condition that $B_1^+ < B_1^-$ or that $B_2^+ < B_2^-$ occurs in some cases [see e.g. low-$T$ part of Fig.~\ref{13Sp4}(b)].

Compared to the sharp changes at $B_1^-$ and $B_2^+$, the $(\tau /B)$ curve shows only a change in the slope at $B_1^+$ and $B_2^-$. 
This asymmetry between $B_1^-$ and $B_1^+$ and between $B_2^+$ and $B_2^-$ may qualitatively be explained within a spirit of the Bean critical state model \cite{BEAN64RMP} (Fig~\ref{Bean}).
When $B_1^+$ is crossed from below to above or $B_2^-$ from above to below, the sample enters a strongly-pinned state.
Since the field gradient built in a weakly-pinned state can be sustained by a large critical current in the strongly-pinned state, the change in the field gradient occurs only gradually from the surface.
Therefore only a bend in the $\tau/B$ curve is observed at $B_1^+$ or $B_2^-$.
On the other hand, when $B_2^+$ is crossed from below to above or $B_1^-$ from above to below, the sample enters a weakly-pinned state from the strongly-pinned one.
Hence the field gradient built in the strongly-pinned state becomes no longer sustainable, and the field gradient quickly changes throughout the sample so that it becomes small enough to be sustained by a small critical current in the weakly-pinned state.
This gives rise to a sudden change in the sample magnetization.

However, if we analyze the minor hysteresis curves in Fig.~\ref{at82deg}(b), it becomes clear that the Bean model quantitatively fails. 
Within the Bean model, a curve branching off from the increasing-field (decreasing-field) curve of the full hysteresis loop at $B_a = B_o$ in the strongly-pinned state is expected to join the decreasing-field (increasing-field) curve of the full loop at $B_a=B_o - 2\Delta B_1$ ($B_0 + 2\Delta B_1$), where $\Delta B_1=\mu_0 J_1 d/2$ is the field difference between the sample surface and center in the strongly-pinned state (Fig.~\ref{Bean}).
We consider the experimental curve branching off from the increasing-field curve at $B_a$ = 23.4 T just below $B_2^+$ in Fig.~\ref{at82deg}(b), which reaches the decreasing-field curve at 19.5 T.
Let us assume $J_1$ = 10$^5$ A/cm$^2$ as $J_c$ of this magnitude has been observed at low fields in doped BaFe$_2$As$_2$ \cite{Tanatar09PRB, Fang11PRB, Demirdis13PRB}.
Since the applied field is roughly parallel to the surface, we might take the sample thickness as $d$: then, $d \sim$ 0.02 mm.
This gives 2$\Delta B_1 \sim$ 0.02 T, too small to explain the observation.
It may be more appropriate to decompose the magnetization and applied field into the $c$-axis and $ab$-plane components.
In the case of the above-mentioned curve, the $c$-axis component of the applied field ($B_a\cos82^{\circ}$) changes from 3.3 to 2.7 T.
The torque is given by $M_cB_a^{ab}-M_{ab}B_a^c$ and is dominated by the first term.
Since the $c$-axis magnetization $M_c$ is caused by the shielding of the $c$-axis component of the field $B_a^c$, the relevant dimension now is the sample length: then $d \sim$ 0.1 mm.
This gives 2$\Delta B_1^c \sim$ 0.1 T for the $c$-axis component, which does not seem sufficient to explain the observation.
We also note in Fig.~\ref{at82deg}(b) the curve branching off from the increasing-field curve at 23.8 T, which is definitely above $B_2^+$ and hence the sample is in the weak-pinning state.
Within the Bean model the curve is expected to approach the decreasing-field curve much more quickly, but it actually goes nearly parallel to the above-discussed curve branching off at $B_a$ = 23.4 T.
Clearly, the behavior of the minor hysteresis loops cannot fully be understood within the Bean model, and it seems necessary to involve complex phase coexistence.

If we assume that the phase below $B_1$ is a Bragg glass, the present phase diagram [Figs. \ref{13Su2}(c) and \ref{13Sp4}(b)] may be interpreted as follows:
The phase between $B_1$ and $B_2$ is a disordered solid phase, which may be a vortex glass \cite{Giamarchi97PRB} or multidomain glass \cite{Menon02PRB}.
The phase above $B_2$ is a vortex liquid, and the irreversibility line is a crossover line separating a pinned and an unpinned liquid.
This interpretation is similar to a proposal in \cite{Banerjee01PhysicaC, Menon02PRB}.
We, however, note the following: those previous works were based on the observation of a single peak of $J_c$ in the peak effect region and associated it with the boundary between the disordered solid and liquid phases.
A very recent small angle neutron scattering study on vanadium, however, claims that the peak effect lies at higher fields and temperatures than the order-disorder transition \cite{Toft-Petersen18NatCommun}.

On the other hand, recent STM studies of the vortex lattice in Co$_{0.0075}$NbSe$_2$ indicate that disordering of a Bragg glass occurs via two phase transitions,  i.e., from the ordered state through an orientational glass where the orientational correlation is maintained to the amorphous vortex glass \cite{Ganguli15SciRep, Ganguli16PRB}.
It is noteworthy that superheating and supercooling effects are observed across either transition.
Two-step disordering has also been reported in a numerical study \cite{Dasgupta03PRL}.
Our $B_1$ and $B_2$ phase transitions might correspond to those two transitions.
It is however to be noted that those studies are for $B \parallel c$.
In the present case, the field is tilted from the $c$ axis.
It may be necessary to consider the two components of $J_c$, i.e., $J_c \parallel c$ and $J_c \bot c$, to explain the existence of the two transitions.

We now consider two other mechanisms that may be related to the anomalous peak effect.
One is a field-induced antiferromagnetism.
Since superconductivity and antiferromagnetism are competing in iron-based superconductors, one might speculate that the latter re-emerges as the former is suppressed by magnetic fields and that it may be related to the anomalous peak effect.
However, the antiferromagnetism is already suppressed before $x$ = 0.3 in Ba$_{1-x}$K$_x$Fe$_2$As$_2$ \cite{Rotter08ACIE}.
$^{75}$As NMR measurements on an $x$ = 0.7 compound indicate that the spin-lattice relaxation rate nearly follows the Korringa relation below $T$ = 100 K, confirming that the composition $x$ = 0.7 is far away from the magnetic instability \cite{Zhang10PRB}.
Further, specific-heat measurements up to $B$ = 13 T for over-doped compositions including $x$ = 0.6, 0.7, and 0.8 show no indication of a field-induced antiferromagnetism \cite{Storey13PRB}.
Therefor this possibility seems unlikely.

The other is the spin paramagnetic effect.
As a magnetic field is applied to a spin-singlet superconductor, the normal-state energy is lowered by spin paramagnetism and a first-order transition to the normal state may occur, or alternatively some theories suggest that a modulated superconducting state, generally called the Fulde-Ferrell-Larkin-Ovchinnikov (FFLO) state, may appear \cite{Maki64Physics1_127, Fulde64PhysRev.135.A550, Larkin64zz, Decroux82Book}.
For a simple BCS superconductor, this critical field (the Pauli limit) is estimated as $B_{po}$ (in Tesla) = 1.84 $T_c$.
In many superconductors, this field is sufficiently larger than the upper critical field, and the spin paramagnetic effect is unimportant.
However, it is not necessarily the case with iron-based superconductors in inplane fields.
For example, KFe$_2$As$_2$ and Ba$_{0.07}$K$_{0.93}$Fe$_2$As$_2$ exhibit a first-order phase transition from the superconducting to normal state for $B \parallel ab$ at low temperatures, and the inplane upper critical field shows an anomalous enhancement, which may be an indication of the FFLO state \cite{Zocco13PRL, Cho17PRL, Terashima13PRB}.
A similar enhancement of $B_{c2} \parallel ab$ at low temperatures is also observed in FeSe \cite{Terashima14PRB}.
In the present case, Figs. \ref{13Su2}(c) and \ref{13Sp4}(b) indicate that the upper critical field for a field direction near $B \parallel ab$ exhibits a saturating trend in an intermediate temperature region as the temperature is lowered, suggesting that the spin paramagnetic effect is important, and then shows an enhancement at still lower temperatures (see the lowest-$T$ point of both figures), which might indicate the FFLO state.
The Pauli limit $B_{po}$ is estimated from $T_c$ to be 35 $\sim$ 39 T for samples 13Su4 (Fig. \ref{13Su4}) and 13Sp4 (Fig. \ref{13Sp4}) and 28 $\sim$ 29 T for sample 13Su2 (Fig. \ref{13Su2}), and the first-order like anomaly at $B_2^+$ appears when $B_2^+$ is close to $B_{po}$.
Since the characteristic field $B_1^-$ for decreasing-field sweeps is fairly away from $B_{po}$, it is not clear whether both of the first-order like anomalies at $B_2^+$ and $B_1^-$ can be explained solely by the spin paramagnetic effect.
Still, a possible role played by the spin paramagnetic effect deserves serious consideration.

Finally, we mention the following feature of the phase diagrams Figs. \ref{13Su2}(c) and \ref{13Sp4}(b):
although the irreversibility field $B_{irr}$ is distinct from the upper critical field $B_{c2}$ at high temperatures, they coincide (within experimental accuracy) as $T$ approaches zero.
This may have implications for an ongoing debate about the exact location of the upper critical field in high-$T_c$ cuprates \cite{Grissonnanche14ncomms, Yu15PRB, Yu16PNAS}.

\section{summary}

We have performed magnetic torque measurements on single crystals of Ba$_{1-x}$K$_x$Fe$_2$As$_2$ ($x \approx$ 0.69 and 0.76).
As the magnetic field is tilted toward the $ab$ plane, the peak effect in torque-vs-field curves becomes pronounced, and it also becomes asymmetric at low temperatures.
Increasing- and decreasing-field curves peak at $B_2^+$ and $B_1^-$, respectively, the former field being markedly higher than the latter at low temperatures.
The increasing- and decreasing-field curves exhibit a sharp change, suggestive of a first-order transition, at the high- and low-field side of the peak, respectively.
Minor hysteresis loops in the peak-effect region exhibit complex history dependence and are difficult to understand with the Bean model.

Defining $B_2^-$ and $B_1^+$ as the counterparts of $B_2^+$ and $B_1^-$, we have constructed the $T$-$B_a$ phase diagram composed of $B_{1}^{+(-)}$, $B_{2}^{+(-)}$, $B_{irr}$, and $B_{c2}$.
If we ascribe the $B_1$ and $B_2$ anomalies to phase transitions of vortex matter, we can suggest two scenarios:
in one scenario, a Bragg glass changes to a disordered solid at $B_1$ and then to a vortex liquid at $B_2$ \cite{Banerjee01PhysicaC, Menon02PRB}, while, in the other, it changes at $B_1$ to an orientational glass where the orientational correlation is maintained and then to an amorphous vortex glass at $B_2$ \cite{Ganguli15SciRep, Ganguli16PRB}.
On the other hand, we note that the first-order-like anomaly is observed when $B_2^+$ is close to the Pauli limit.
This may indicate that the spin paramagnetic effect plays some role in causing the anomalous peak effect.
So far the peak effect under a strong influence of the spin paramagnetic effect has not seriously been studied experimentally nor theoretically and hence deserves further studies.

\section{acknowledgments}
We are grateful to the late Professor James S. Brooks (NHMFL, Florida State University) for his continuous support during this work.
This work was supported by the Transformative Research-Project on Iron Pnictides (TRIP) from JST and also by JSPS KAKENHI Grant Numbers JP26400373 and JP17K05556, and Scientific Research B (No. 24340090).
A portion of this work was performed at the National High Magnetic Field Laboratory, which is supported by National Science Foundation Cooperative Agreement No. DMR-1157490 and the State of Florida.



%

\end{document}